\documentclass[runningheads]{llncs}
\usepackage[T1]{fontenc}
\usepackage{graphicx}
\usepackage{amsmath}
\usepackage{amssymb}
\usepackage{mathtools}
\usepackage{graphicx}
\usepackage{graphics}
\usepackage{svg}
\usepackage{rotating}
\usepackage{subcaption}

\usepackage{cite}

\usepackage{pifont}

\usepackage[colorinlistoftodos]{todonotes}
\usepackage[colorlinks=true, allcolors=blue]{hyperref}
\usepackage{caption}
\DeclareCaptionFont{mysize}{\fontsize{9}{9}\selectfont}
\usepackage[title]{appendix}
\usepackage{amsfonts}
\usepackage{array,multirow}
\usepackage{mwe}
\usepackage{booktabs}
\usepackage[ruled,vlined,algo2e]{algorithm2e}
\usepackage{xcolor,colortbl}
\definecolor{rred}{rgb}{0.95,0.4,0.3}
\definecolor{bblue}{rgb}{0.4,0.6,1.00}

\newcommand\Tstrut{\rule{0pt}{2.4ex}}         
\newcommand\Bstrut{\rule[-0.9ex]{0pt}{0pt}}   

\usepackage{tabularx,booktabs}

\begin{document}
\title{Meta-information-aware Dual-path Transformer for Differential Diagnosis of Multi-type Pancreatic Lesions in Multi-phase CT}
\titlerunning{Differential Diagnosis of Multi-type Pancreatic Lesions in Multi-phase CT}
%
\author{Bo Zhou\inst{1,2}\thanks{This work was supported by Alibaba Group through Alibaba Research Intern Program.} \and Yingda Xia\inst{1} \and Jiawen Yao\inst{1} \and Le Lu\inst{1} \and Jingren Zhou\inst{1} \and Chi Liu\inst{2,3} \and James S. Duncan\inst{2,3,4} \and Ling Zhang\inst{1}}

\institute{DAMO Academy, Alibaba Group
\and
Department of Biomedical Engineering, Yale University
\and
Department of Radiology and Biomedical Imaging, Yale University
\and
Department of Electrical Engineering, Yale University
}
\authorrunning{B. Zhou et al.}
%
%
\maketitle              
%
\begin{abstract}
Pancreatic cancer is one of the leading causes of cancer-related death. Accurate detection, segmentation, and differential diagnosis of the full taxonomy of pancreatic lesions, i.e., normal, seven major types of lesions, and ``other'' lesions, is critical to aid the clinical decision-making of patient management and treatment. However, existing work focus on segmentation and classification for very specific lesion types (PDAC) or groups. Moreover, none of the previous work considers using lesion prevalence-related non-imaging patient information to assist the differential diagnosis. To this end, we develop a meta-information-aware dual-path transformer and exploit the feasibility of classification and segmentation of the full taxonomy of pancreatic lesions. Specifically, the proposed method consists of a CNN-based segmentation path (S-path) and a transformer-based classification path (C-path). The S-path focuses on initial feature extraction by semantic segmentation using a UNet-based network. The C-path utilizes both the extracted features and meta-information for patient-level classification based on stacks of dual-path transformer blocks that enhance the modeling of global contextual information. A large-scale multi-phase CT dataset of 3,096 patients with the pathology-confirmed pancreatic lesion class labels, voxel-wise manual annotations of lesions from radiologists, and patient meta-information, was collected for training and evaluations. Our results show that our method can enable accurate classification and segmentation of the full taxonomy of pancreatic lesions, approaching the accuracy of the radiologist's report and significantly outperforming previous baselines. Results also show that adding the common meta-information, i.e., gender and age, can boost the model's performance, thus demonstrating the importance of meta-information for aiding pancreatic disease diagnosis. 

\keywords{Pancreatic Lesion \and Dual-path Transformer \and Meta-information Aware \and Differential Diagnosis.}
\end{abstract}

\section{Introduction}
\vspace{-0.18cm}
Pancreatic cancer is the third leading cause of death among all cancers in the United States, and has the poorest prognosis among all solid malignancies with a 5-year survival rate of about 10\% \cite{grossberg2020multidisciplinary}. Early diagnosis and treatment are crucial, which can potentially increase the 5-year survival rate to about 50\% \cite{conroy2018folfirinox}. In clinical practice, pancreatic patient management is based on the pancreatic lesion type and the potential of the lesion to become invasive cancer. However, pancreatic lesions are often hard to reach by biopsy needle because of the deep location in the abdomen and the complex structure of surrounding organs and vessels. To this end, accurate imaging-based differential diagnosis of pancreatic lesion type is critical to aid the clinical decision-making of patient management and treatment, e.g., surgery, monitoring, or discharge \cite{zhao20213d,springer2019multimodality}. Multi-phase Computed Tomography (CT) is the first-line imaging tool for pancreatic disease diagnosis. However, accurate differential diagnosis of pancreatic lesions is very challenging because 1) the same type of lesion may have different textures, shapes, and contrast patterns across multi-phase CT, and 2) pancreatic ductal adenocarcinoma (PDAC) accounts for the majority of cases, e.g., $>$60\%, in pathology-confirmed patient population, leading to a long-tail problem. 

Most related work in automatic pancreatic CT image analysis focus on segmentation of certain types of pancreatic lesions, e.g., PDAC and pancreatic neuroendocrine tumor (PNET). UNet-based detection-by-segmentation approaches have been extensively studied for the detection of PDAC \cite{xia2020detecting,zhang2020robust,zhou2019hyper,zhou2017deep,zhu2019multi} and PNET \cite{zhu2021segmentation}. Shape-induced information, e.g., tubular structure of dilated duct, is exploited to improve the PDAC detection \cite{liu2019joint,wang2020deep}. Graph-based classification network is proposed for pancreatic patient risk stratification and management \cite{zhao20213d}. There are also recent attempts in the detection and classification of PDAC and nonPDAC using non-contrast CT \cite{xia2021effective}. However, none of the previous work has yet attempted to address the key clinical need for detection and classification of full taxonomy of pancreatic lesions, i.e., PDAC, PNET, solid pseudopapillary tumor (SPT), intraductal papillary mucinous lesion (IPMN), mucinous cystic lesion (MCN), chronic pancreatitis (CP), serous cystic lesion (SCN) \cite{springer2019multimodality}, and other rare types that can be further classified into other benign and other malignant. Furthermore, no methods consider adding lesion prevalence-related non-imaging patient information to aid the diagnosis. For example, based on epidemiological data, the incidence of MCN, SCN, and SPT in women is significantly higher than that in men, and MCN, SCN, and SPT has a higher prevalence in young-age, middle-age, and old-age female, respectively \cite{hu2022cystic}. Integrating easily-accessible clinical patient meta-information, e.g., gender and age in the DICOM head, as classification feature inputs could potentially further improve the diagnosis accuracy without needing radiologists' manual input. 

To address these challenges and unmet needs, we propose a meta-information-aware dual-path transformer (MDPFormer) for classification and segmentation of the full taxonomy of pancreatic lesions, including normal, seven major types of pancreatic lesions, other malignant, and other benign. Motivated by the recent dual-path design of Mask Transformers~\cite{wang2021max, cheng2021per}, the proposed MDPFormer consists of a segmentation path (S-path) and a classification path (C-path). The S-path focuses on initial feature extraction by semantic segmentation (normal, PDAC, and nonPDAC) using a CNN-based network. Then, the C-path utilizes both meta-information and the extracted features for individual-level classification (normal, PDAC, PNET, SPT, IPMN, MCN, CP, SCN, other benign, and other malignant) based on stacked dual-path transformer blocks that enhance the modeling of global contextual information. We curated a large-scale multi-phase CT dataset with the pathology-confirmed pancreatic lesion class labels, voxel-wise manual annotations of lesions from radiologists, and patient meta-information. To our knowledge, this model is the most comprehensive to date, and is trained on a labeled dataset (2,372 patients' multi-phase CT scans) larger than that used in previous studies \cite{xia2022felix,zhao20213d,park2022deep}. We independently test our method on a test set consisting of one whole year of 724 consecutive patients with pancreatic lesions from a high-volume pancreatic cancer center. The experimental results show that our method enables accurate classification and segmentation of the full taxonomy of pancreatic lesions, approaching the accuracy of radiologists' reports (by second-line senior readers via referring to current and previous imaging, patient history, and clinical meta-information). Our method without meta-information input demonstrates superior classification and segmentation performance as compared to previous baselines. Adding the meta-information-aware design further boosts the model's performance, demonstrating the importance of meta-information for improving pancreatic disease diagnosis. 

\begin{figure*}[t]
\centering
\includegraphics[width=0.92\textwidth]{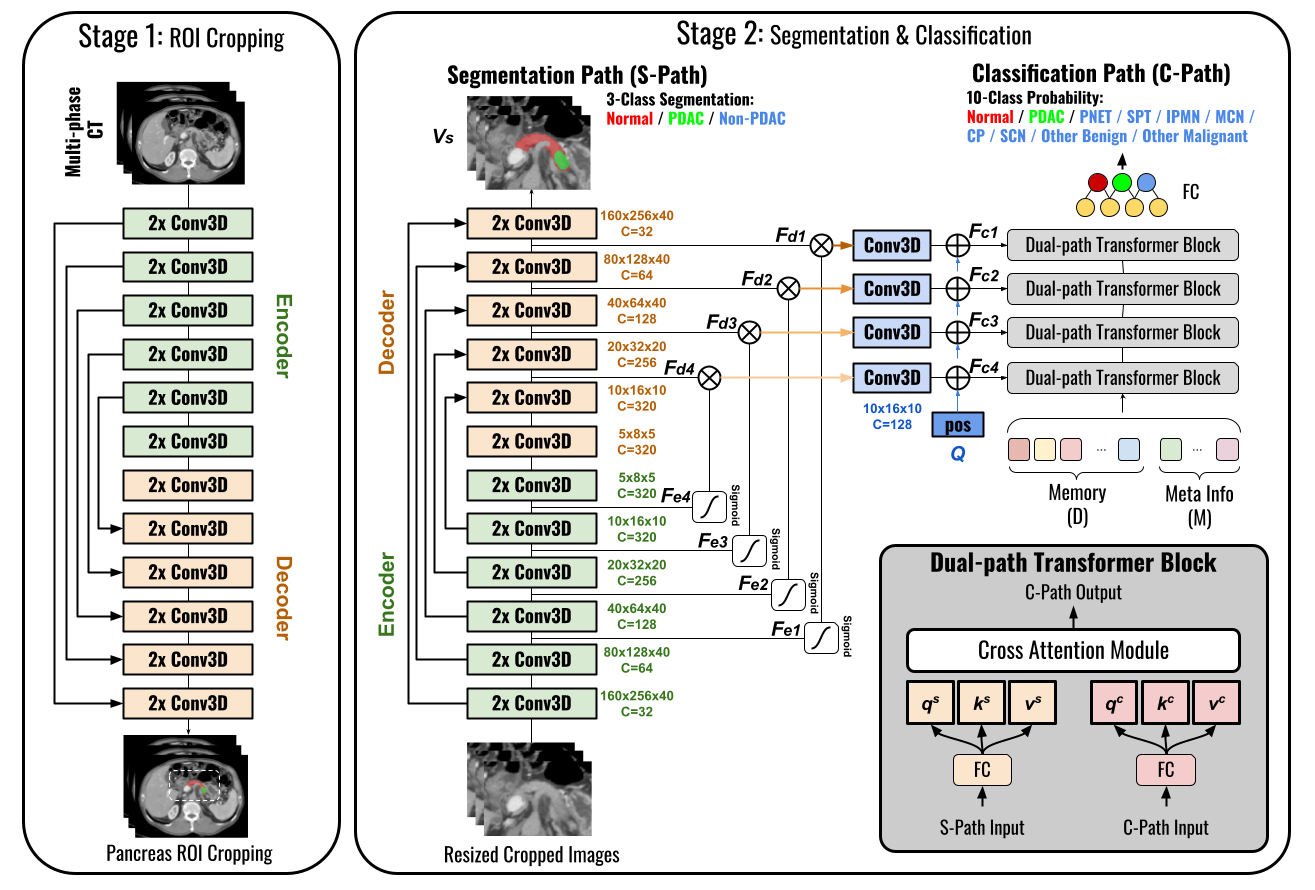}
\caption{The overall pipeline and the detailed structure of our MDPFormer. In stage 1, the pancreas sub-volume is cropped based on a coarse pancreas segmentation mask. In stage 2, the resized pancreas sub-volume is inputted into the MDPFormer for segmentation (left path) and classification (right path). The design of dual-path transformer block in the classification path is illustrated on the bottom right (grey box).}
\label{fig:network} \vspace{-3mm}
\end{figure*}

\vspace{-0.18cm}
\section{Methods}
\vspace{-0.18cm}
The general pipeline of our method is illustrated in Figure \ref{fig:network}. Our pipeline consists of two stages. In the first stage, we use a localization UNet \cite{cciccek20163d} to segment out the pancreas from the whole CT volume. The sub-volume containing the pancreas is then cropped out based on the segmentation mask. In the second stage, the resized sub-volume is inputted into the meta-information-aware dual-path transformer (MDPFormer) to segment and classify the pancreatic lesions. Details are elaborated in the following sections.

\vspace{0.1cm}
\noindent\textbf{Meta-information-aware Dual-path Transformer.} For classification, we denote $\mathcal{H}_c = \{0,1,2,\cdots,9\}$ for the ten patient/lesion classes, i.e., normal, PDAC, PNET, SPT, IPMN, MCN, CP, SCN, other benign, and other malignant. For segmentation, we group the last eight classes into nonPDAC and denote $\mathcal{H}_s = \{0,1,2\}$ for the grouped three patient classes, i.e., normal, PDAC, and nonPDAC. The goal is to enable a more balanced initial class distribution for segmentation, while enabling feature extraction for the full pancreatic lesion taxonomy classification. The training set is thus formulated as $S = \{ (X_i, M_i, Y_i, Z_i) | i=1,2,\cdots,N \}$, where $X_i$ is the cropped pancreas sub-volume of the i-th patients, $M_i$ is the patient meta information (gender and age), $Y_i \in \mathcal{H}_s$ is the 3-class voxel-wise annotation with the same spatial size as $X_i$, and $Z_i \in \mathcal{H}_c$ is the 10-class volume-wise label that confirmed by pathology or clinical records. 

The MDPFormer consists of two paths, including a segmentation path (S-Path) and a classification path (C-Path). The goal of S-path is to extract rich feature representations of the lesion and pancreas at multiple scales by first segmenting the image into three general classes. Given a input $X$ and a segmentation network $G_s$, we have
\begin{equation}
    V_s, F_{d1}, F_{d2}, F_{d3}, F_{d4}, F_{e1}, F_{e2}, F_{e3}, F_{e4} = G_s(X)
\end{equation}
where $V_s$ is the segmentation output, $F_{d1}, F_{d2}, F_{d3}, F_{d4}$ are the multi-scale features from the decoder, $F_{e1}, F_{e2}, F_{e3}, F_{e4}$ are the multi-scale features from the encoder. Here, we deploy a 3D UNet \cite{cciccek20163d} as the S-Path backbone network. Instead of directly using the decoder features as C-path input, we combine the multi-scale encoder and decoder features by 
\begin{equation}
    F_c = f_c (F_d * \sigma(F_e)) + Q
\end{equation}
where $\sigma$ is the sigmoid function for generating attention from the encoder features to guide decoder feature outputs, $f_c$ is a convolution layer that further refines the S-Path feature output, and $Q$ is the learnable position embedding feature that provides position representation to aid the transformer in C-path. $F_c$ is the extracted feature from the S-Path which is used for C-Path input.

The C-Path consists of four consecutive dual-path transformer blocks, where each block takes both the S-Path feature and the global memory feature as inputs. Denote $D$ as the initial 1D memory feature which is randomly initialized learnable parameters~\cite{wang2021max}, we fuse the patient meta-information with the initial memory feature by 
\begin{equation}
    F_m = [D, M]
\end{equation}
where 
$D$ and $M$ are concatenated in the length dimension and $M$ is the meta-information, i.e., patient gender and age, in this work. In each block, we use a cross-attention module to fuse $F_m$ and $F_c$. First, we compute S-Path queries $q^s$, keys $k^s$, and values $v^p$, by learnable linear projections of the S-Path feature $F_s$ at each feature location. Similarly, queries $q^c$, keys $k^c$, and values $v^c$ are computed from C-path global memory feature $F_c$ with another set of projection matrices. The cross-attention output can then be calculated as follows:
\begin{equation}
    y^c = softmax (q^c \cdot k^{cs}) v^{cs} ,
\end{equation}
\begin{equation}
    k^{cs} = \begin{bmatrix} k^c \\ k^s \end{bmatrix}, v^{cs} = \begin{bmatrix} v^c \\ v^s \end{bmatrix} ,
\end{equation}
where $[\cdot]$ is the concatenation operator in the channel dimension to fuse the values and keys from both paths. The output $y^c$ is then inputted into the next block as the $F_m$ memory feature input. Using the C-path feature output from the last dual-path transformer block, we predict the final classification $P$ with two fully connected layers and a softmax. The overall training objective can thus be formulated as:
\begin{equation}
    \mathcal{L}_{all} = \mathcal{L}_s (V_s, Y) + \mathcal{L}_c (P, Z)
\end{equation}
where $\mathcal{L}_s (\cdot)$ is the Dice loss function for segmentation training, and $\mathcal{L}_c (\cdot)$ is the cross-entropy loss for classification training. 

\vspace{-0.18cm}
\section{Experimental Results}
\vspace{-0.18cm}
{\bf Data Preparation.} We collected a large-scale multi-phase CT dataset consisting of 3,096 patients from a high-volume pancreatic cancer institution. Each multi-phase CT consists of noncontrast, arterial, and venous phase CT. The data were consecutively collected from 2015-2020. All the 724 patients scanned during 2020 were used as the independent test set, and the rest of the 2,372 patients scanned from 2015-2019 were used as the training set. The training set includes 707 normal, 1,088 PDAC, 110 PNET, 68 SPT, 162 IPMN, 32 MCN, 64 CP, 93 SCN, 48 other benign, and 24 other malignant cases. The test set includes 202 normal, 283 PDAC, 34 PNET, 25 SPT, 73 IPMN, 9 MCN, 29 CP, 38 SCN, 14 other benign, and 17 other malignant cases. All patients with lesions were confirmed by surgical pathology, while normal patients were confirmed by radiology reports and at least 2-year follow-ups. The annotation of lesions was performed collaboratively by an experienced radiologist (with 14 years of specialized experience in pancreatic imaging) and an auto-segmentation model on either arterial or venous phase CT, whichever with better lesion visibility. More specifically, the radiologist first annotates some data to train an auto-segmentation model to segment the remaining data, which is then checked/edited by the radiologist. The CT phases were registered using DEEDS \cite{heinrich2013mrf}. The gender and age information were extracted from the DICOM head as meta information inputs. The gender is converted to a binary value, i.e., 0 for female and 1 for male. The age is normalized between 0-1 by dividing the value by 100. 

\vspace{0.05cm}
\noindent{\bf Implementation Details.} All CT volumes were resampled into $0.68 \times 0.68 \times 3.0$ mm spacing and normalized into zero mean and unit variance. In the training phase of MDPFomer, we cropped the foreground 3D bounding box of the pancreas region, randomly pad small margins on each dimension, and resized the sub-volume into $160 \times 256 \times 40$ (Y$\times$X$\times$Z) for input. We deployed a 5-fold cross-validation strategy using the 2,372 training set to train and validate five models. During inference, the five models' predictions were ensemble by averaging the prediction results. For each fold, we first pre-trained the S-path network for 1000 epochs, and then trained the whole model in an end-to-end fashion with an SGD optimizer. The initial learning rate was set to $1 \times 10^{-3}$ with cosine decay, and the batch size was set to 3. The localization UNet in the first stage followed the same training protocol. 

\vspace{0.05cm}
\noindent{\bf Compared Methods and Evaluation Metrics.} Our method is compared with two types of baseline approaches. One is the ``segmentation for classification (S4C)'' method where a segmentation network, i.e., nnUNet \cite{isensee2021nnu} or (nn)UNETR \cite{hatamizadeh2022swin,isensee2021nnu}, is first deployed for semantic segmentation of the ten classes on the cropped sub-volume. We then classify the patient based on the class-wise lesion segmentation size. Specifically, if one or multiple lesion classes were presented in the segmentation, we classify the patient to the lesion class with the largest segmentation size; Otherwise, we classify the patient as normal. Note that we implement UNETR \cite{hatamizadeh2022swin} in the nnUNet framework \cite{isensee2021nnu}, called (nn)UNETR, which shows substantially better results than the original UNETR implementation on our data. The other baseline is the CNN-based segmentation-to-classification method. We use the exact same structure of S-path in MDPFormer, and extract all encoder and decoder multi-scale features. Then, we apply global max pooling on each feature map, concatenate them and forward them into two fully connected layers for classification. We also compared our performance with the radiology report, which represents the clinical read performance of second-line senior radiologists (via referring to current and previous imaging, patient history, and clinical information) in the high-volume pancreatic cancer center. The classification performance was evaluated by class-wise accuracy, regular accuracy, and balanced accuracy. The confusion matrices were also reported for detailed evaluation. The segmentation performance was evaluated by the Dice coefficient or score on each class of pancreatic lesion or normal.

\begin{table} [t]
\scriptsize
\centering
\caption{Evaluation of classification performance on lesion diagnosis (\%). Both averaged accuracy (second last row) and balanced accuracy (last row) are reported.}
\label{tab:cls}
    \begin{tabular}{>{\centering\arraybackslash}p{1.8cm}|>{\centering\arraybackslash}p{1.4cm}|>{\centering\arraybackslash}p{1.8cm}||>{\centering\arraybackslash}p{1.6cm}|>{\centering\arraybackslash}p{1.6cm}|>{\centering\arraybackslash}p{1.8cm}||>{\centering\arraybackslash}p{1.2cm}}
        \hline 
        \textit{\textbf{CLASSIFY}}   & \textbf{nnUNet} & \textbf{(nn)UNETR} & \textbf{SPath+FC} & \textbf{DPFormer} & \textbf{MDPFormer} & \textbf{Report} \Tstrut\Bstrut\\
        \hline
        \textbf{Normal}     & 96.0    & 96.2 & 97.0 & \textbf{\textcolor{blue}{99.0}} & \textbf{\textcolor{red}{99.5}} & 100       \Tstrut\Bstrut\\
        \hline
        \textbf{PDAC}       & \textbf{\textcolor{blue}{94.3}}    & 94.1 & \textbf{\textcolor{blue}{94.3}} & \textbf{\textcolor{blue}{94.3}} & \textbf{\textcolor{red}{96.5}} & 93.3       \Tstrut\Bstrut\\
        \hline
        \textbf{PNET}       & 38.2    & 37.5 & 35.3 & \textbf{\textcolor{blue}{47.1}} & \textbf{\textcolor{red}{47.1}} & 70.6       \Tstrut\Bstrut\\
        \hline
        \textbf{SPT}        & 64.0    & 62.8 & 60.0 & \textbf{\textcolor{blue}{64.0}} & \textbf{\textcolor{red}{72.0}} & 84.0       \Tstrut\Bstrut\\
        \hline
        \textbf{IPMN}       & \textbf{\textcolor{red}{69.9}}    & \textbf{\textcolor{blue}{68.1}} & 43.8 & 60.3 & 65.8 & 68.5       \Tstrut\Bstrut\\
        \hline
        \textbf{MCN}        & 0.0     & 0.0  & \textbf{\textcolor{blue}{11.1}} & \textbf{\textcolor{blue}{11.1}} & \textbf{\textcolor{red}{33.3}} & 33.3       \Tstrut\Bstrut\\
        \hline
        \textbf{CP}         & 6.9     & 17.2 & 24.1 & \textbf{\textcolor{blue}{31.0}} & \textbf{\textcolor{red}{44.8}} & 69.0       \Tstrut\Bstrut\\
        \hline
        \textbf{SCN}        & 44.7    & 42.1 & \textbf{\textcolor{blue}{50.0}} & \textbf{\textcolor{blue}{50.0}} & \textbf{\textcolor{red}{55.3}} & 42.1       \Tstrut\Bstrut\\
        \hline
        \textbf{Other-BEN}  & 0.0     & 0.0  & 21.4 & \textbf{\textcolor{blue}{28.6}} & \textbf{\textcolor{red}{35.7}} & 35.7       \Tstrut\Bstrut\\
        \hline
        \textbf{Other-MLG}  & 0.0     & 0.0  & 0.0  & \textbf{\textcolor{blue}{11.7}} & \textbf{\textcolor{red}{11.7}} & 17.6       \Tstrut\Bstrut\\
        \hline\hline
        \underline{\textbf{Regular Acc}}    & 77.4    & 77.4 & 76.2 & \textbf{\textcolor{blue}{79.8}} & \textbf{\textcolor{red}{82.9}} & 84.0      \Tstrut\Bstrut\\
        \hline
        \underline{\textbf{Balance Acc}}    & 41.4    & 41.8 & 43.6 & \textbf{\textcolor{blue}{49.7}} & \textbf{\textcolor{red}{56.2}} & 61.4       \Tstrut\Bstrut\\
        
        \hline
    \end{tabular}
\end{table}

\vspace{-0.22cm}
\subsection{Main Results}
\vspace{-0.12cm}
The classification results are summarised in Table \ref{tab:cls}. Comparing DPFormer without meta-information-aware to the previous baseline methods, i.e., UNet-based S4C, UNETR-based S4C, and S-Path+FC, we can see that DPFormer can already outperform all the baselines in 9 out of 10 classes and achieve the highest balanced accuracy of $49.71\%$. In general, it is challenging to use the conventional segmentation approaches to directly segment out the 10 classes and perform classification based on it. For S4C approaches, we can see the classification accuracy of MCN, CP, other benign, and other malignant are all zero or near zero. While S-Path+FC provide slightly better classification result with the additional FC layer for classification, DPFormer with dual path transformer and better feature fusion provides better results. With the meta-information-aware design that incorporates additional gender and age information, our MDPFormer utilizes those easily-accessible tumor-type-related non-imaging information, thus achieving further improved classification balanced accuracy of 56.17\%. 

\begin{figure*}[htb!]
\centering
\includegraphics[width=1.00\textwidth]{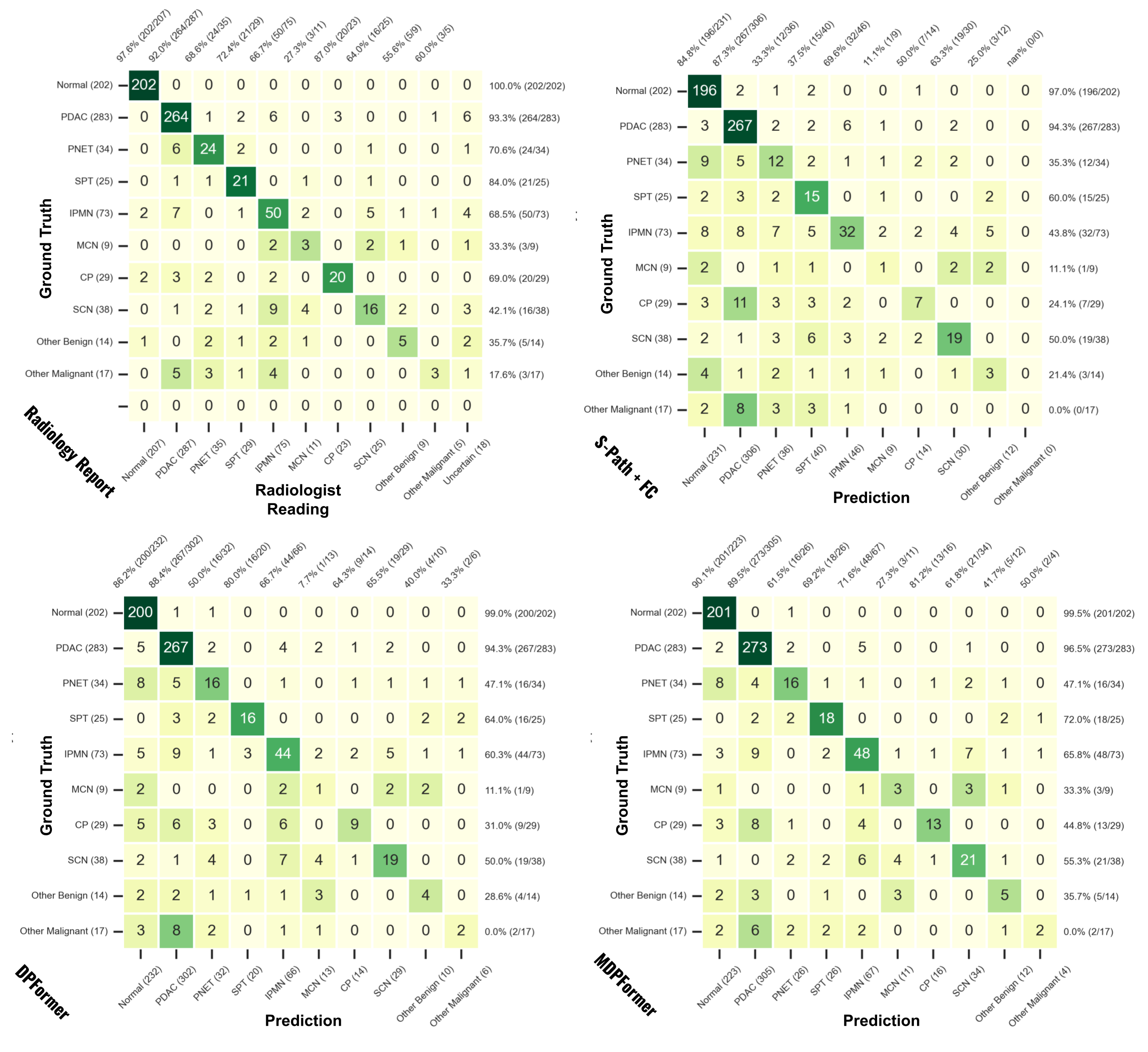}
\caption{Comparison of classification performance using confusion matrices.}
\label{fig:cm} \vspace{-6mm}
\end{figure*}

\begin{figure*}[htb!]
\centering
\includegraphics[width=0.94\textwidth]{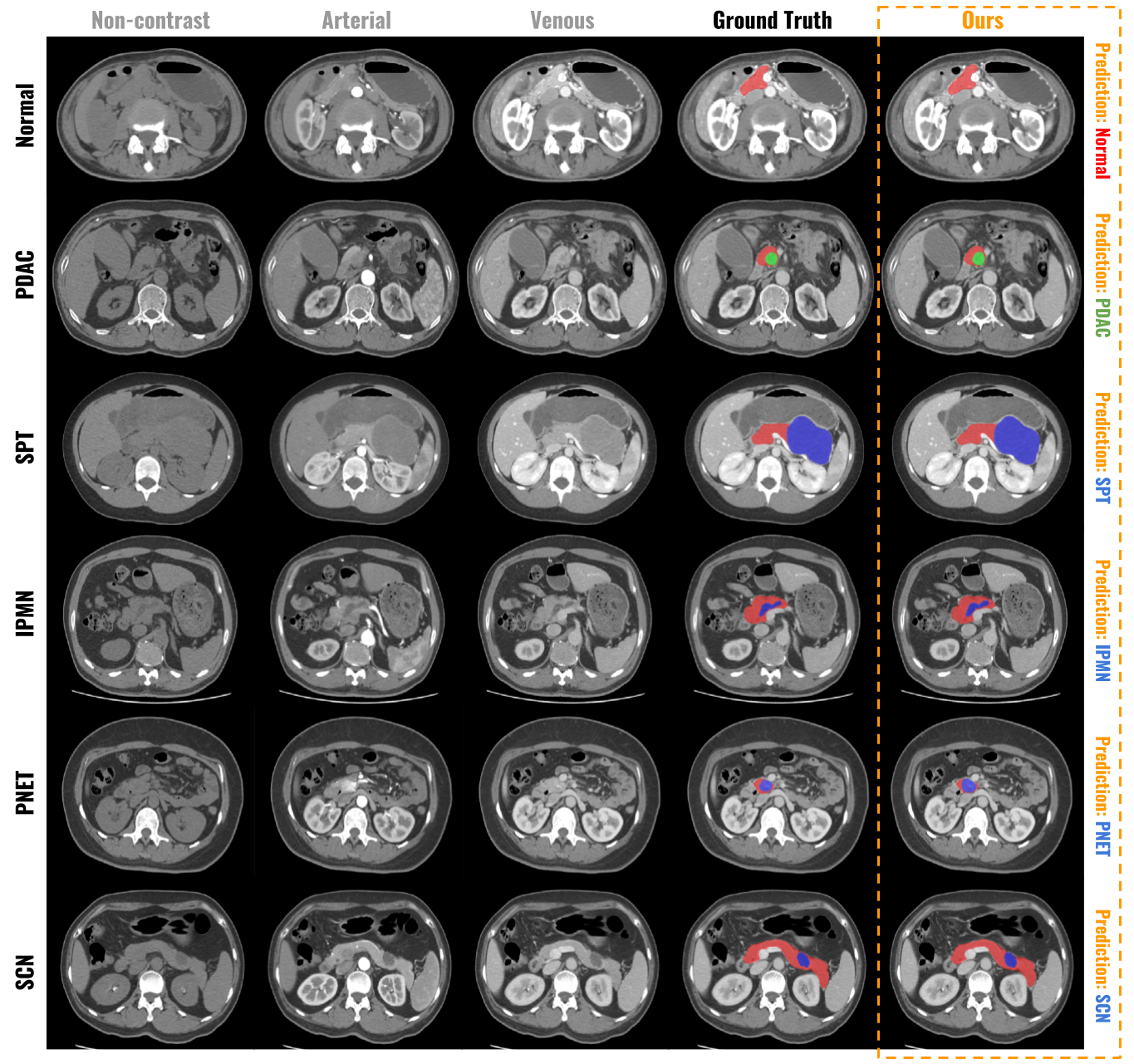}
\caption{Examples of classification and segmentation outputs from our MDPFormer. Ground truth lesion classes are annotated on the left and the predicted classes are shown on the right. Segmented pancreas is depicted in Red; lesion in Green or Blue.}
\label{fig:results} 
\end{figure*}

The classification results compared to the radiology report are also shown in Figure \ref{tab:cls} and elaborated in Figure \ref{fig:cm}. The balanced classification accuracy of the radiology report is 61.41\%. Adding meta-information improves our method's balanced classification performance from 49.71\% to 56.17\%, approaching the performance of the radiology report. Our method also provides better PDAC (96.5\% vs. 93.3\%) and SCN (55.4\% vs. 42.1\%) diagnosis accuracy as compared to the reports, which is critical since PDAC is of the highest priority among all pancreatic abnormalities with a 5-year survival rate of approximately 10\% and is the most common type (>60\% of all pathology-confirmed pancreatic lesions). In general, the radiology reports that perform diagnosis with more meta-information, e.g., patient history, tumor markers, previous report, etc., provide better classification accuracy. Thus, adding additional meta-information may further improve our method's performance. In addition, unlike radiology reports that only give the final diagnosis, our MPDFormer provides both classification probabilities and class-wise lesion segmentation outputs with explainability. Examples of our MDPFormer's classification and segmentation results are shown in Figure \ref{fig:results}.

The accuracy of the ``Report" for the normal class is 100\% (Table \ref{tab:cls} and Figure \ref{fig:cm}). This is because our normal cases were selected based on the radiology reports reporting an absence of pancreatic lesions. Actually, the radiologists' specificity for the normal pancreas is 93\%--96\% in a pancreas CT interpretation setting \cite{park2022deep}. Our MDPFormer has a higher specificity (99.5\%) than radiologists, making it a reliable detection tool for pancreatic lesions in practice.

Ablative studies for the segmentation performance are summarized in Table \ref{tab:seg}. For MDPFormer, DPFormer, and SPath+FC, please note that the nonPDAC segmentation class is assigned by the final classification prediction. Similar to the observation from classification evaluations, it is difficult for nnUNet and UNETR to directly perform 10-class segmentation with averaged Dice scores of 0.360 and 0.373 reported, respectively. On the other hand, our MDPFormer can provide significantly better segmentation performances for all 10 normal and lesion classes ($p < 0.001$) and achieve an averaged Dice score of $0.604$. Comparing MDPFormer to DPFormer, we can also see that adding the meta-information improves the segmentation performance (averaged Dice of 0.604 versus 0.502). Note that the Dice scores reported in Table \ref{tab:seg} are generally higher than that reported in previous studies \cite{zhao20213d,xia2022felix,zhang2020robust}. This is mainly because the ground truth annotations are generated semi-automatically. Nevertheless, the above results clearly demonstrate the superiority of our MDPFormer over compared methods.

\begin{table} [t]
\scriptsize
\centering
\caption{Evaluation of segmentation performance on normal and lesion (Dice).}
\label{tab:seg}
    \begin{tabular}{>{\centering\arraybackslash}p{1.9cm}|>{\centering\arraybackslash}p{1.7cm}|>{\centering\arraybackslash}p{1.7cm}||>{\centering\arraybackslash}p{1.7cm}|>{\centering\arraybackslash}p{1.7cm}|>{\centering\arraybackslash}p{1.9cm}}
        \hline 
        \textit{\textbf{SEGMENT}}   & \textbf{nnUNet} & \textbf{(nn)UNETR} & \textbf{SPath+FC} & \textbf{DPFormer} & \textbf{MDPFormer} \Tstrut\Bstrut\\
        \hline
        \textbf{Normal}     & 0.950$\pm$0.118    & 0.940$\pm$0.109 & 0.951$\pm$0.107 & \textbf{\textcolor{blue}{0.953$\pm$0.096}} & \textbf{\textcolor{red}{0.958$\pm$0.069}}        \Tstrut\Bstrut\\
        \hline
        \textbf{PDAC}       & 0.863$\pm$0.157    & 0.860$\pm$0.149 & \textbf{\textcolor{blue}{0.866$\pm$0.189}} & 0.865$\pm$0.199 & \textbf{\textcolor{red}{0.869$\pm$0.196}}        \Tstrut\Bstrut\\
        \hline
        \textbf{PNET}       & 0.259$\pm$0.302    & 0.288$\pm$0.310 & 0.352$\pm$0.381 & \textbf{\textcolor{blue}{0.355$\pm$0.391}} & \textbf{\textcolor{red}{0.456$\pm$0.390}}        \Tstrut\Bstrut\\
        \hline
        \textbf{SPT}        & 0.513$\pm$0.370    & 0.537$\pm$0.352 & 0.624$\pm$0.326 & \textbf{\textcolor{blue}{0.662$\pm$0.429}} & \textbf{\textcolor{red}{0.766$\pm$0.414}}        \Tstrut\Bstrut\\
        \hline
        \textbf{IPMN}       & 0.475$\pm$0.304    & 0.468$\pm$0.302 & 0.515$\pm$0.340 & \textbf{\textcolor{blue}{0.518$\pm$0.390}} & \textbf{\textcolor{red}{0.598$\pm$0.382}}        \Tstrut\Bstrut\\
        \hline
        \textbf{MCN}        & 0.071$\pm$0.159    & 0.098$\pm$0.189 & 0.211$\pm$0.446 & \textbf{\textcolor{blue}{0.312$\pm$0.395}} & \textbf{\textcolor{red}{0.416$\pm$0.441}}        \Tstrut\Bstrut\\
        \hline
        \textbf{CP}         & 0.051$\pm$0.098    & 0.112$\pm$0.253 & 0.280$\pm$0.323 & \textbf{\textcolor{blue}{0.349$\pm$0.338}} & \textbf{\textcolor{red}{0.382$\pm$0.335}}        \Tstrut\Bstrut\\
        \hline
        \textbf{SCN}        & 0.431$\pm$0.351    & 0.428$\pm$0.348 & 0.484$\pm$0.303 & \textbf{\textcolor{blue}{0.587$\pm$0.441}} & \textbf{\textcolor{red}{0.765$\pm$0.438}}        \Tstrut\Bstrut\\
        \hline
        \textbf{Other-BEN}  & 0.0$\pm$0.0        & 0.0$\pm$0.0     & 0.227$\pm$0.397 & \textbf{\textcolor{blue}{0.293$\pm$0.364}} & \textbf{\textcolor{red}{0.459$\pm$0.422}}        \Tstrut\Bstrut\\
        \hline
        \textbf{Other-MLG}  & 0.0$\pm$0.0        & 0.0$\pm$0.0     & 0.088$\pm$0.247 & \textbf{\textcolor{blue}{0.129$\pm$0.284}} & \textbf{\textcolor{red}{0.373$\pm$0.394}}        \Tstrut\Bstrut\\
        \hline\hline
        \underline{\textbf{Average}}    & 0.361    & 0.373 & 0.464 & \textbf{\textcolor{blue}{0.502}} & \textbf{\textcolor{red}{0.604}}       \Tstrut\Bstrut\\
        
        \hline
    \end{tabular}
\end{table}

\begin{figure*}[htb!]
\centering
\includegraphics[width=0.96\textwidth]{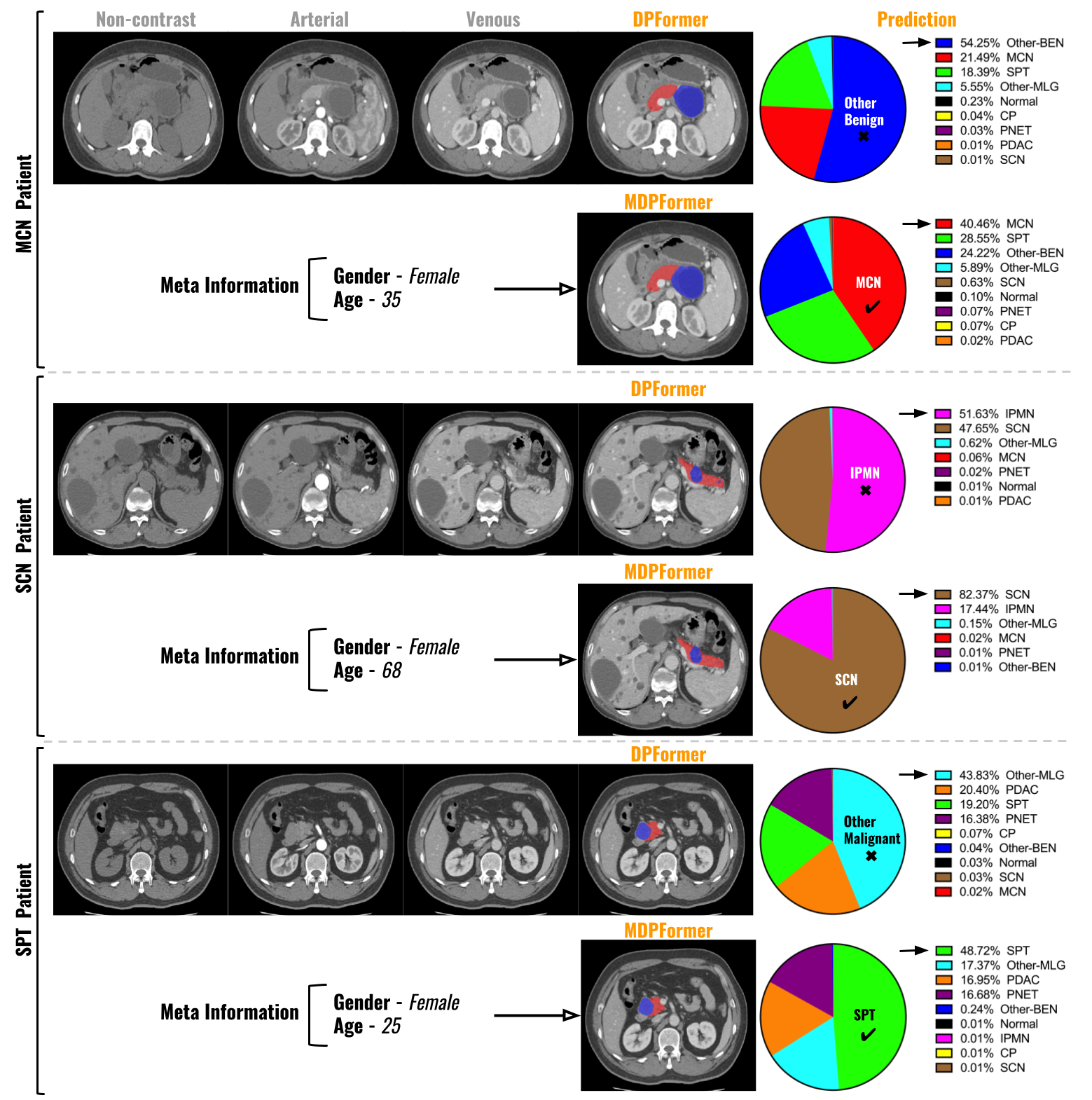}
\caption{Case studies of three patients with MCN, SCN, and SPT. The classification probability predictions of DPFormer and MDPFormer models are shown on the right. }
\label{fig:abla} \vspace{-5mm}
\end{figure*}

Next, we provide three patient case studies to show the impact of adding meta-information for classifying the pancreas lesion. The studies are illustrated in Figure \ref{fig:abla}, including three patients with MCN, SCN, and SPT, respectively. Using DPFormer without patient meta-information, the MCN, SCN, and SPT were misclassified as other benign, IPMN, and other malignant, respectively. The MDFormer adding the gender and age information to the imaging information provide more accurate tumor probability predictions. For example, for the female 68-year-old patient with SCN, the maximal probability predicted by DPFormer is 51.63\% for IPMN, while MDPFormer with meta-information provides the maximal probability of 82.37\% for SCN. 

\vspace{-0.12cm}
\subsection{Discussion}
\vspace{-0.22cm}
In this work, we present a meta-information-aware dual-path transformer (MDPFormer) for the classification and segmentation of pancreatic lesions in multi-phase CT. The MDFormer consists of an S-path and C-path, where the S-path focuses on initial feature extraction by group-level segmentation and the C-path utilizes both meta-information and the extracted features for individual-level classification. Compared to previous baselines, our method without meta-information input already shows superior classification and segmentation performance. Adding the meta-information-aware design further boost these performances, demonstrating the importance of meta-information when diagnosing specific pancreatic lesion type. Our MDPFormer is an open framework with several key components adjustable, which could potentially further improve our future performances. First, we used a simple UNet architecture with two consecutive convolution layers at each scale level for feature extraction. Using more advanced segmentation network blocks maybe can provide richer feature representations for better classification and segmentation performances. Second, we only used meta-information of patient gender and age as inputs, which can be automatically extracted from every DICOM data in practice. Adding additional non-imaging information, e.g., family history, symptoms (weight loss, jaundice), and other patient records (CA 19-9 blood test), may further potentially improve MDPFormer to better match the performance of the radiologists who have access to those non-imaging information for diagnosis. Those are important research directions for our future work.

\vspace{-0.18cm}
\section{Conclusion}
\vspace{-0.28cm}
This paper presents a new meta-information-aware dual-path transformer for classification and segmentation of the full taxonomy of pancreatic lesions. Our experimental results show that the proposed dual-path transformer can efficiently incorporate the patient meta-information and the extracted image features from the CNN-based segmentation path to make accurate pancreatic lesion classification and segmentation. We demonstrate that our method achieves better performance than previous baselines and approaches the accuracy of radiology reports. Our system could be a useful assistant tool for pancreatic lesion detection, segmentation, and diagnosis in the clinical reading environment.

%
%
%
\bibliographystyle{splncs04}
\bibliography{bibliography}

\end{document}